\documentclass[aps,pra,superscriptaddress,twocolumn,floatfix,a4paper]{revtex4}

\usepackage{graphicx,graphics,epsfig}   
\usepackage{dcolumn}    
\usepackage{bm}         
\usepackage{amsmath}    
\usepackage{verbatim}   
\usepackage{color}      
\usepackage{subfigure}  
\usepackage{times,natbib}
\usepackage{amsmath,amsfonts,amssymb,graphics,graphics,color,times}

\usepackage{latexsym}
\usepackage{amsmath}
\usepackage{amssymb}
\usepackage{amsfonts}
\usepackage{amsthm}
\usepackage{mathrsfs}
\usepackage{color,verbatim,graphics}
\usepackage{psfrag}
\DeclareMathAlphabet{\mathrsfs}{U}{rsfs}{m}{n}
\DeclareMathAlphabet{\mathpzc}{OT1}{pzc}{m}{it}
\DeclareMathAlphabet{\matheus}{U}{eus}{m}{n}
\DeclareMathAlphabet{\mathbbold}{U}{bbold}{m}{n}

\def\one{\leavevmode\hbox{\small1\normalsize\kern-.33em1}}

\newcommand{\ba}{\begin{eqnarray}}
\newcommand{\ea}{\end{eqnarray}}
\newcommand{\ban}{\begin{eqnarray*}}
\newcommand{\ean}{\end{eqnarray*}}

\begin{document}

\title{More non-locality in the three-qubit Greenberger-Horne-Zeilinger state}

\author{Tam\'as V\'ertesi}
\affiliation{Institute of Nuclear Research of the Hungarian Academy of Sciences
H-4001 Debrecen, P.O. Box 51, Hungary}

\author{K\'aroly F. P\'al}
\affiliation{Institute of Nuclear Research of the Hungarian Academy of Sciences
H-4001 Debrecen, P.O. Box 51, Hungary}

\date{\today}


\begin{abstract}
The non-local properties of the noisy three-qubit
Greenberger-Horne-Zeilinger (GHZ) states,
$\rho_v=v|GHZ\rangle\langle GHZ|+(1-v)\one/8$ parameterized by the
visibility $0\le v\le 1$ are investigated. Based on the violation
of the $2\times 2\times 2$-setting Mermin inequality, $\rho_v$ is
non-local for the parameter range $1/2<v\le 1$. It has been posed
whether additional settings would allow to lower the threshold
visibility. Here we report on Bell inequalities giving a threshold
value smaller than $v=1/2$. This rules out the possibility of a
local hidden variable model in the limit of $v=1/2$. In
particular, the lowest threshold visibility we found is $v=0.496057$,
attainable with $5\times 5\times 5$ settings, whereas the most
economical one in number of settings corresponds to $3\times
3\times 4$ settings. The method which enabled us to obtain these
results, and in particular the about 10000 tight Bell inequalities
giving $v<1/2$ are also discussed in detail.
\end{abstract}

\maketitle

\section{Introduction}

Quantum theory allows correlations between remote systems, which
are fundamentally different from classical correlations. Quantum
entanglement is in the heart of this phenomenon \cite{horo}. In
particular, separated observers may carry out local measurements
on entangled quantum states on such a way that the correlations
they generate are outside the set of common cause correlations.
This is known as Bell's theorem \cite{bell}.

Moreover, such quantum correlations may find application in novel
device-independent information tasks. They enable perfect security
\cite{ekert,secure} or randomness generation \cite{randomness}
without the need to trust the internal working of the devices.

It is desirable to find states which tolerate the largest noise in
order these protocols be useful. In this regard, let us note a
link between the strength of violation of certain multipartite
Bell inequalities and the security of quantum communication
protocols \cite{SG}.

For the simplest case of a Werner state \cite{werner}, which is
the maximally entangled singlet state mixed with white noise, the
CHSH \cite{CHSH} seems to be the best inequality for a modest
number of settings ($\le10$) \cite{zuk99,gruca}, giving the
critical visibility $v=1/\sqrt 2$. Recently, this value has been
slightly overcome with $465$ number of settings \cite{ver08}.

When we turn to a maximally entangled state shared by more than
two parties, the threshold visibility drops down rapidly
\cite{mermin} (exponentially in the number of qubits). Even for
three parties by sharing the noisy 3-qubit GHZ state the critical
visibility becomes $v_{crit}=1/2$ using the Mermin inequality
(note that the separability limit corresponds to $v_{sep}=1/5$
\cite{DCT}). To the best of our knowledge, there has been no known
Bell inequality giving a lower threshold. Let us note that in
recent studies \cite{kz,gruca} using a sophisticated numerical
method the lowest value found up to settings $5\times 5\times 5$,
matched $v=1/2$ associated with the Mermin inequality. In the
present paper, we provide a systematic study to explore Bell
inequalities with violations beyond the Mermin value. Our findings 
may also bear relevance to the problem of
simulating GHZ-correlations with classical communication
\cite{simul,BG}.

We start by introducing the Mermin inequality
(Sec.~\ref{sec:mermin}). Then, in Sec.~\ref{sec:method} a
heuristic method is given to generate tight Bell inequalities with
potentially large violations. In Sec.~\ref{sec:results}, we
present the results by listing the Bell inequalities according to
different aspects. In Sec.~\ref{sec:analytic}, a symmetric one is
highlighted, with small coefficients, and a closed solution is
given for the maximum quantum violation yielding a critical
visibility lower than $1/2$. The paper ends in Sec.~\ref{sec:conc}
by summarizing the results and posing open questions.

\section{Mermin inequality as a case study}\label{sec:mermin}

Let us represent the $2\times 2\times 2$ setting Mermin
polynomial~\cite{mermin} in terms of three-party correlators,
\begin{align}
M\equiv& A_0B_0C_0-A_0B_1C_1-A_1B_0C_1-A_1B_1C_0,
\label{eq:Mermin}
\end{align}
where the above polynomial is to be understood as a sum of
expectation values; for example, $A_1B_1C_0$ denotes the
average value of the product of the outcomes of Alice, Bob and
Cecil when both Alice and Bob performs the second measurement and
Cecil performs the first one. The local bound of this inequality
is ${\cal L}=2$.

For a given number of measurement settings and outcomes the
correlations which can be described by a local realistic model
define a polytope (the so-called Bell polytope), which is a convex
set with a finite number of extreme points. Bell inequalities
define the limits on these correlations. The Mermin inequality is
a facet of the $2\times 2\times 2$ local 3-party-correlation
polytope \cite{WW} (involving only full-tripartite correlation
terms) and also for the $2\times 2\times 2$ Bell polytope
\cite{sliwa} (involving single party marginals and 2-party
correlators as well). Note that in order to get the lowest
critical visibility for a given number of settings and outcomes,
it is always sufficient to consider the facets of the local
polytope. We refer to inequalities associated with these facets as
tight Bell inequalities (hence Mermin inequality is a tight one).
Furthermore, it suffices to focus on inequivalent Bell
inequalities, that is, on those which are not equivalent under
relabelling of measurement settings, outcomes and exchange of
parties. We would also like to mention that the inequivalent Bell
inequalities to be presented turn out to be tight both in the
full-tripartite-correlation space and in the full probability
space.

Let us now take the noisy 3-qubit GHZ state,
\begin{equation}
\rho_v=v|GHZ\rangle\langle GHZ|+(1-v)\frac{\one}{8},
\label{eq:noisyGHZ}
\end{equation}
where $|GHZ\rangle=\left(|000\rangle+|111\rangle\right)/\sqrt 2$
is the 3-qubit GHZ state~\cite{GHZ}, and $v$ is the visibility
parameter. With equatorial von Neumann measurements, corresponding
to traceless observables, the maximum quantum value saturates the
algebraic limit (${\cal Q}=4$). This is the no-signalling limit as
well, hence Mermin inequality exhibits pseudotelepathy \cite{BBT}.
Note, that the inequalities we present in the following will not
have this property, that is, the no-signalling bound will never be
saturated. We denote the threshold visibility for which quantum
correlations turn to classical by $v_{crit}$ (namely, for
$v>v_{crit}$ in Eq.~(\ref{eq:noisyGHZ}) there do not exist local
realistic correlations reproducing the quantum ones). From the
form of Eq.~(\ref{eq:noisyGHZ}) it follows the ratio
$v_{crit}={\cal L}/{\cal Q}$, where $\cal L$ is the local bound
and $\cal Q$ is the quantum bound achievable with traceless
observables and yields $v_{crit}=1/2$ for the 3-qubit GHZ state
based on the Mermin inequality. In the following we will exhibit
several inequalities which go beyond this bound.

\begin{table*}[tbm]
 \caption{Bell inequalities with critical visibility $v_{crit}<0.5$. The two cases with three measurement
 settings for two parties and one for one party, and the best ten cases with three measurement settings
 for two parties and one for the other two. Symbols ${\cal L}$ and $M_{ijk}$ denote
 the local bound and the Bell coefficients, respectively.}
 \vskip 0.2truecm
 \centering
 \begin{tabular}{l l l r r r r r r r r r r r r r r r r}
 \hline\hline
 Case&$v_{crit}$&${\cal L}$&&&&&&&&&$M_{ijk}$\\
 \hline
 $V_{343}^{1}$&0.49967&44&\quad 4&-4&0&-10&\quad 3&4&4&-1&\quad 3&-6&14&9\\
 &&&\quad -4&-3&-4&-1&\quad 3&-3&0&-2&\quad -3&-4&-4&1\\
 &&&\quad -6&3&14&-9&\quad -4&-3&-4&-1&\quad -4&4&0&10\\
 \hline
 $V_{343}^{2}$&0.49972&24&\quad -2&5&3&0&\quad 4&5&-1&-8&\quad -2&0&-2&-2\\
 &&&\quad 2&5&-3&8&\quad -2&5&3&0&\quad 2&0&2&2\\
 &&&\quad 2&0&2&2&\quad -2&0&-2&-2&\quad -2&0&2&0\\
 \hline
 $V_{344}^{1}$&0.49851&16&\quad 0&2&1&-1&\quad -1&-1&-1&1&\quad -1&-1&1&1&\quad 0&0&-1&1\\
 &&&\quad -3&-5&-1&-3&\quad -4&2&-1&1&\quad -1&1&0&-2&\quad 0&-2&2&0\\
 &&&\quad 3&-1&-2&-2&\quad 3&-3&0&0&\quad 0&2&1&-1&\quad 0&-2&1&1\\
 \hline
 $V_{344}^{2}$&0.49860&76&\quad -25&14&-11&-6&\quad 6&1&-7&6&\quad 13&12&4&-1&\quad 10&-3&-6&-7\\
 &&&\quad 0&14&11&7&\quad -7&0&9&-6&\quad 15&11&3&-1&\quad -10&-3&5&8\\
 &&&\quad -7&0&6&-3&\quad 3&-1&-6&-6&\quad 0&-1&-1&2&\quad -6&0&5&-5\\
 \hline
 $V_{344}^{3}$&0.49863&24&\quad -2&3&-2&-1&\quad -3&1&2&0&\quad 5&1&0&-4&\quad 0&3&2&-5\\
 &&&\quad 2&-1&-2&1&\quad -2&1&-2&-1&\quad 0&0&0&0&\quad -2&2&0&0\\
 &&&\quad -2&2&-2&0&\quad -3&2&2&-1&\quad -5&-1&0&4&\quad 8&3&2&3\\
 \hline
 $V_{344}^{4}$&0.49875&20&\quad -2&3&1&-2&\quad 2&1&-1&-2&\quad 0&-4&0&-4&\quad 2&6&0&0\\
 &&&\quad 2&-2&1&1&\quad -2&-2&-1&3&\quad 0&-4&0&-4&\quad -2&0&0&-6\\
 &&&\quad -2&-1&0&1&\quad -2&1&0&-1&\quad 0&0&0&0&\quad 0&2&0&-2\\
 \hline
 $V_{344}^{5}$&0.49876&20&\quad 0&-1&2&-1&\quad -2&-1&-1&0&\quad 0&1&0&-1&\quad -2&-1&-1&0\\
 &&&\quad 2&2&2&0&\quad 4&-1&1&4&\quad -2&1&-1&0&\quad 0&-2&2&-4\\
 &&&\quad 2&1&2&1&\quad -2&0&2&-4&\quad -2&2&1&1&\quad -6&-1&3&4\\
 \hline
 $V_{344}^{6}$&0.49879&20&\quad -2&-2&-1&-1&\quad 0&-1&1&2&\quad 2&1&0&1&\quad 0&0&0&0\\
 &&&\quad 3&-2&2&1&\quad 2&1&-1&2&\quad 6&1&4&-3&\quad -3&0&3&0\\
 &&&\quad -3&2&1&-2&\quad -2&-2&0&-2&\quad 0&-2&4&2&\quad -3&0&3&0\\
 \hline
 $V_{344}^{7}$&0.49881&16&\quad 0&-2&1&-1&\quad 0&-2&-1&1&\quad 3&1&0&2&\quad 3&-1&4&0\\
 &&&\quad -1&-1&3&-1&\quad 0&-1&-1&2&\quad -2&2&5&1&\quad -3&0&-3&0\\
 &&&\quad 1&-1&-2&0&\quad 0&1&0&1&\quad -1&1&1&-1&\quad 0&-1&1&0\\
 \hline
 $V_{344}^{8}$&0.49883&20&\quad 1&-1&-1&-1&\quad -1&1&0&-2&\quad 1&-2&-1&0&\quad 1&0&0&1\\
 &&&\quad 1&1&3&-1&\quad -1&3&0&2&\quad -2&6&-2&-2&\quad 4&4&-1&1\\
 &&&\quad 0&2&2&-2&\quad 0&2&2&2&\quad 3&0&-3&-2&\quad -3&-4&1&0\\
 \hline
 $V_{344}^{9}$&0.49886&40&\quad 1&5&-3&-3&\quad 0&-4&-4&4&\quad -8&0&-4&-4&\quad -7&-7&1&1\\
 &&&\quad -1&3&2&-2&\quad -2&0&-2&-4&\quad 0&4&2&-2&\quad 1&-1&-2&0\\
 &&&\quad 2&-6&5&1&\quad 2&4&2&-4&\quad -8&12&2&6&\quad -8&-6&1&-1\\
 \hline
 $V_{344}^{10}$&0.49891&20&\quad -4&-4&0&0&\quad -1&2&-1&-2&\quad -5&0&-1&-2&\quad -2&2&0&2\\
 &&&\quad 0&0&-1&1&\quad 1&-1&0&-2&\quad 2&-2&0&0&\quad -1&1&-1&-1\\
 &&&\quad 4&4&1&-1&\quad -2&3&1&-2&\quad 1&-6&-1&-2&\quad -1&1&-1&1\\
 \hline
 \end{tabular}
 \label{table:443}
 \end{table*}

\begin{table*}[tbm]
 \caption{The best ten Bell inequalities with four measurement settings per party with critical visibility
 $v_{crit}<0.5$. Symbols ${\cal L}$ and $M_{ijk}$ denote
 the local bound and the Bell coefficients, respectively.}
 \vskip 0.2truecm
 \centering
 \begin{tabular}{l l l r r r r r r r r r r r r r r r r}
 \hline\hline
 Case&$v_{crit}$&${\cal L}$&&&&&&&&&$M_{ijk}$\\
 \hline
 $V_{444}^{1}$&0.49699&12&\quad 0&-1&1&0&\quad -2&-3&-2&-1&\quad 0&0&0&0&\quad 2&0&-1&-1\\
 &&&\quad 0&-1&1&0&\quad 2&-1&-1&0&\quad 0&0&0&0&\quad -2&2&0&0\\
 &&&\quad 1&1&1&1&\quad 0&-1&0&1&\quad -1&1&0&0&\quad 0&-1&1&0\\
 &&&\quad 1&1&-1&-1&\quad 0&1&-1&0&\quad -1&1&0&0&\quad 0&1&-2&1\\
 \hline
 $V_{444}^{2}$&0.49720&12&\quad 0&0&0&0&\quad 0&-1&1&0&\quad 0&-1&1&0&\quad 0&0&0&0\\
 &&&\quad 0&1&2&1&\quad 2&0&0&2&\quad 2&-1&-1&0&\quad 0&0&-1&1\\
 &&&\quad 0&1&1&0&\quad 0&-2&1&-1&\quad 0&2&-1&1&\quad 0&1&1&0\\
 &&&\quad 0&0&-1&-1&\quad 2&-1&0&1&\quad 2&2&1&-3&\quad 0&-1&0&-1\\
 \hline
 $V_{444}^{3}$&0.49751&12&\quad 1&2&-2&-1&\quad 0&0&1&-1&\quad 1&-1&1&1&\quad 0&1&0&1\\
 &&&\quad 0&-1&0&-1&\quad 0&0&0&0&\quad 0&-1&-2&1&\quad 0&2&2&0\\
 &&&\quad 0&-2&1&-1&\quad 0&-1&1&0&\quad 1&-2&0&1&\quad 1&-1&-2&0\\
 &&&\quad 1&1&1&1&\quad 0&-1&0&1&\quad 0&0&-1&1&\quad -1&0&0&1\\
 \hline
 $V_{444}^{4}$&0.49765&12&\quad -2&-2&-1&-1&\quad -2&-1&0&1&\quad -1&0&1&0&\quad -1&1&0&0\\
 &&&\quad -1&-1&1&-1&\quad 2&0&1&1&\quad 1&-1&1&1&\quad 0&0&1&-1\\
 &&&\quad -1&-1&1&1&\quad 1&0&-1&-2&\quad 1&0&0&1&\quad -1&1&0&0\\
 &&&\quad 0&0&-1&1&\quad -1&1&0&0&\quad 1&-1&0&0&\quad 0&0&1&-1\\
 \hline
 $V_{444}^{5}$&0.49773&24&\quad -4&-3&-2&-1&\quad -3&-1&0&4&\quad -2&3&-1&0&\quad -1&-1&1&-1\\
 &&&\quad -3&0&-1&2&\quad -1&3&4&0&\quad 3&-2&0&1&\quad -1&1&-1&-1\\
 &&&\quad -2&-1&2&1&\quad 0&4&1&-1&\quad -1&0&2&-3&\quad 1&-1&1&1\\
 &&&\quad -1&2&1&2&\quad 4&0&-1&-1&\quad 0&1&-3&2&\quad -1&-1&1&-1\\
 \hline
 $V_{444}^{6}$&0.49785&16&\quad -3&-3&-2&-2&\quad -2&-1&-1&2&\quad -2&2&0&0&\quad -1&0&1&0\\
 &&&\quad -2&1&-1&0&\quad -1&2&1&0&\quad 1&-2&2&-1&\quad 0&-1&0&1\\
 &&&\quad -1&-1&0&2&\quad 0&2&0&-2&\quad -2&1&1&0&\quad 1&0&-1&0\\
 &&&\quad 0&-1&1&0&\quad 1&-1&0&0&\quad -1&-1&-1&-1&\quad 0&-1&0&1\\
 \hline
 $V_{444}^{7}$&0.49786&24&\quad -3&-3&-2&-2&\quad -2&3&-2&-1&\quad -1&-2&0&3&\quad 0&-2&0&-2\\
 &&&\quad -3&-1&-2&2&\quad 3&0&1&2&\quad -2&1&3&0&\quad -2&2&0&0\\
 &&&\quad -2&-2&1&1&\quad -2&1&0&-3&\quad 0&3&1&-2&\quad 0&0&0&0\\
 &&&\quad -2&2&1&3&\quad -1&2&-3&2&\quad 3&0&-2&-1&\quad -2&0&0&-2\\
 \hline
 $V_{444}^{8}$&0.49795&24&\quad -1&2&1&-2&\quad 2&-2&0&2&\quad -1&0&-1&0&\quad 2&2&0&0\\
 &&&\quad -2&0&0&-6&\quad 1&2&-1&-2&\quad -2&-4&0&-2&\quad -1&2&1&-2\\
 &&&\quad -1&-4&0&-3&\quad 1&-1&1&-1&\quad -1&3&0&4&\quad -1&0&-1&0\\
 &&&\quad -2&-2&1&3&\quad -2&3&0&1&\quad -2&1&-1&2&\quad -2&2&0&-2\\
 \hline
 $V_{444}^{9}$&0.49803&24&\quad -2&-4&-2&0&\quad 3&2&-1&2&\quad -1&2&2&-1&\quad 0&0&-1&1\\
 &&&\quad 0&-2&-1&-1&\quad -1&1&0&0&\quad -2&0&-1&-1&\quad 1&-1&0&0\\
 &&&\quad 2&-6&-1&1&\quad -2&1&-4&-1&\quad 2&-2&-2&-2&\quad 2&1&1&-2\\
 &&&\quad 0&4&-4&0&\quad 0&4&-3&-1&\quad -3&0&1&0&\quad 1&0&0&-3\\
 \hline
 $V_{444}^{10}$&0.49806&20&\quad 0&2&-1&-1&\quad -1&-1&0&0&\quad 0&1&1&2&\quad 1&-2&-2&-3\\
 &&&\quad 0&-2&-1&-1&\quad -1&-1&0&0&\quad 0&1&-3&-2&\quad 1&-2&2&1\\
 &&&\quad 0&0&2&2&\quad 0&0&2&-2&\quad 0&-2&-4&-2&\quad 0&2&0&-6\\
 &&&\quad 0&0&0&0&\quad 0&0&2&-2&\quad 0&0&-2&2&\quad 0&0&-4&4\\
 \hline
 \end{tabular}
 \label{table:444}
 \end{table*}

 \begin{table*}[tbm]
 \caption{Some special Bell inequalities with four measurement settings per party with critical visibility
 $v_{crit}<0.5$. For two cases the absolute value of the Bell coefficients are no more than one,
 while three cases are symmetric for all participants.
 Symbols ${\cal L}$ and $M_{ijk}$ denote
 the local bound and the Bell coefficients, respectively.}
 \vskip 0.2truecm
 \centering
 \begin{tabular}{l l l r r r r r r r r r r r r r r r r}
 \hline\hline
 Case&$v_{crit}$&${\cal L}$&&&&&&&&&$M_{}$\\
 \hline
 $V_{444}^{U1}$&0.49890&8&\quad -1&-1&-1&-1&\quad -1&-1&-1&1&\quad -1&0&0&-1&\quad -1&0&0&1\\
 &&&\quad -1&0&0&-1&\quad 1&-1&0&0&\quad 0&1&0&1&\quad 0&0&0&0\\
 &&&\quad 0&-1&-1&0&\quad 0&1&0&1&\quad -1&0&1&0&\quad 1&0&0&-1\\
 &&&\quad 0&0&0&0&\quad 0&-1&1&0&\quad 0&-1&1&0&\quad 0&0&0&0\\
 \hline
 $V_{444}^{U2}$&0.49955&8&\quad -1&-1&-1&-1&\quad -1&0&0&1&\quad -1&0&1&0&\quad -1&1&0&0\\
 &&&\quad -1&0&0&1&\quad 0&0&1&1&\quad 1&0&1&0&\quad 0&0&0&0\\
 &&&\quad 0&-1&0&1&\quad 1&0&1&0&\quad 0&0&-1&-1&\quad -1&1&0&0\\
 &&&\quad 0&0&-1&1&\quad 0&0&0&0&\quad 0&0&-1&1&\quad 0&0&0&0\\
 \hline
 $V_{444}^{S1}$&0.49895&128&\quad -22&-10&-3&-1&\quad -10&4&-1&-13&\quad -3&-1&11&-1&\quad -1&-13&-1&9\\
 &&&\quad -10&4&-1&-13&\quad 4&10&13&1&\quad -1&13&-12&-4&\quad -13&1&-4&12\\
 &&&\quad -3&-1&11&-1&\quad -1&13&-12&-4&\quad 11&-12&-9&-12&\quad -1&-4&-12&-15\\
 &&&\quad -1&-13&-1&9&\quad -13&1&-4&12&\quad -1&-4&-12&-15&\quad 9&12&-15&16\\
 \hline
 $V_{444}^{S2}$&0.49903&12&\quad -3&-1&0&0&\quad -1&0&-1&0&\quad 0&-1&1&0&\quad 0&0&0&2\\
 &&&\quad -1&0&-1&0&\quad 0&1&0&-1&\quad -1&0&1&0&\quad 0&-1&0&-1\\
 &&&\quad 0&-1&1&0&\quad -1&0&1&0&\quad 1&1&1&1&\quad 0&0&1&-1\\
 &&&\quad 0&0&0&2&\quad 0&-1&0&-1&\quad 0&0&1&-1&\quad 2&-1&-1&0\\
 \hline
 $V_{444}^{S3}$&0.49990&400&\quad -62&-53&-43&-30&\quad -53&34&-21&-8&\quad -43&-21&-1&35&\quad -30&-8&35&-7\\
 &&&\quad -53&34&-21&-8&\quad 34&-15&19&-26&\quad -21&19&44&-16&\quad -8&-26&-16&-12\\
 &&&\quad -43&-21&-1&35&\quad -21&19&44&-16&\quad -1&44&-11&-6&\quad 35&-16&-6&17\\
 &&&\quad -30&-8&35&-7&\quad -8&-26&-16&-12&\quad 35&-16&-6&17&\quad -7&-12&17&44\\
 \hline
 \end{tabular}
 \label{table:444spec}
 \end{table*}

 \begin{table*}[tbm]
 \caption{Examples of Bell inequalities with critical visibility
 $v_{crit}<0.5$ with five measurement settings for at least one party.
 Symbols ${\cal L}$ and $M_{ijk}$ denote
 the local bound and the Bell coefficients, respectively.}
 \vskip 0.2truecm
 \centering
 \begin{tabular}{l l l r r r r r r r r r r r r r r r r r r r r r r r r r}
 \hline\hline
 Case&$v_{crit}$&${\cal L}$&&&&&&&&&&&$M_{ijk}$\\
 \hline
 $V_{555}^{1}$&0.496057&24&\quad 2&-2&1&-1&2&\quad 0&2&0&-2&0&\quad 0&-2&0&2&0&\quad -2&-1&1&0&-2&\quad 0&1&0&1&0\\
 &&&\quad 0&1&0&2&1&\quad -2&0&0&0&2&\quad -2&0&0&0&-2&\quad 0&2&2&4&0&\quad 0&1&-2&2&-1\\
 &&&\quad 0&2&2&0&0&\quad 0&0&0&0&0&\quad 0&0&0&0&0&\quad 0&0&2&-2&0&\quad 0&2&0&2&0\\
 &&&\quad -2&0&0&0&2&\quad 0&-2&0&2&0&\quad 0&-2&0&2&0&\quad 2&0&0&0&-2&\quad 0&0&0&0&0\\
 &&&\quad 0&1&-1&1&1&\quad 2&0&0&0&-2&\quad 2&0&0&0&2&\quad 0&1&1&2&0&\quad 0&0&-2&1&-1\\
 \hline
 $V_{555}^{3}$&0.496081&12&\quad 0&0&0&0&0&\quad 0&-1&1&0&0&\quad 1&2&1&0&0&\quad 0&-1&-1&0&0&\quad -1&2&1&0&0\\
 &&&\quad 0&0&0&0&0&\quad 0&2&-2&0&0&\quad 1&1&0&0&0&\quad 0&0&0&0&0&\quad -1&1&-2&0&0\\
 &&&\quad 0&0&0&0&0&\quad 0&1&-1&0&0&\quad 0&0&0&0&0&\quad 0&0&0&0&0&\quad 0&-1&1&0&0\\
 &&&\quad 0&0&1&0&-1&\quad 0&0&0&0&0&\quad -1&1&1&1&0&\quad 1&0&1&1&1&\quad 0&-1&-1&0&0\\
 &&&\quad 0&0&-1&0&1&\quad 0&0&0&0&0&\quad -1&0&0&1&0&\quad 1&-1&-2&1&-1&\quad 0&-1&-1&0&0\\
 \hline
 $V_{555}^{S1}$&0.496485&12&\quad 0&0&-1&-1&0&\quad 0&0&0&0&0&\quad -1&0&0&0&1&\quad -1&0&0&0&1&\quad 0&0&1&1&0\\
 &&&\quad 0&0&0&0&0&\quad 0&0&1&-1&0&\quad 0&1&-1&0&0&\quad 0&-1&0&1&0&\quad 0&0&0&0&0\\
 &&&\quad -1&0&0&0&1&\quad 0&1&-1&0&0&\quad 0&-1&-2&1&0&\quad 0&0&1&-1&0&\quad 1&0&0&0&1\\
 &&&\quad -1&0&0&0&1&\quad 0&-1&0&1&0&\quad 0&0&1&-1&0&\quad 0&1&-1&2&0&\quad 1&0&0&0&1\\
 &&&\quad 0&0&1&1&0&\quad 0&0&0&0&0&\quad 1&0&0&0&1&\quad 1&0&0&0&1&\quad 0&0&1&1&0\\
 \hline
 $V_{555}^{U1}$&0.4976723&16&\quad 1&0&0&1&0&\quad 0&1&-1&-1&1&\quad 0&0&0&1&1&\quad 1&0&-1&1&-1&\quad 0&1&0&0&-1\\
 &&&\quad 0&0&0&1&1&\quad 0&0&-1&1&0&\quad 0&0&-1&0&-1&\quad 0&1&1&-1&-1&\quad 0&1&-1&-1&1\\
 &&&\quad 0&1&-1&0&0&\quad -1&0&0&-1&0&\quad 1&1&0&0&0&\quad 0&0&1&0&1&\quad 0&0&0&-1&-1\\
 &&&\quad 0&-1&0&-1&0&\quad -1&1&0&0&0&\quad 1&0&0&1&0&\quad 0&1&-1&-1&1&\quad 0&-1&-1&1&1\\
 &&&\quad -1&0&-1&1&-1&\quad 0&0&0&-1&-1&\quad 0&1&-1&0&0&\quad -1&0&0&-1&0&\quad 0&1&0&1&0\\
 \hline
 $V_{555}^{U2}$&0.497977&16&\quad 0&0&0&0&0&\quad 1&0&-1&1&1&\quad 0&-1&0&0&1&\quad 0&1&1&-1&1&\quad -1&0&0&0&1\\
 &&&\quad 0&1&0&1&0&\quad -1&1&0&0&0&\quad 0&-1&1&0&0&\quad -1&0&0&0&1&\quad 0&1&-1&-1&-1\\
 &&&\quad 0&0&1&0&1&\quad 0&0&-1&0&1&\quad -1&-1&-1&-1&0&\quad 0&0&1&0&-1&\quad -1&1&0&-1&-1\\
 &&&\quad 0&1&0&1&0&\quad -1&0&-1&0&0&\quad 1&1&1&0&1&\quad 0&1&0&1&0&\quad 0&-1&0&0&-1\\
 &&&\quad 0&0&-1&0&-1&\quad 1&1&-1&-1&0&\quad 0&0&-1&1&0&\quad -1&0&0&0&1&\quad 0&-1&-1&0&0\\
 \hline
 $V_{455}^{1}$&0.496062&12&\quad -1&0&1&0&0&\quad 0&0&-1&0&1&\quad 0&2&2&0&0&\quad 0&0&-1&0&-1&\quad -1&-2&1&0&0\\
 &&&\quad -1&0&1&0&0&\quad 0&0&-1&0&1&\quad 0&-2&-1&0&-1&\quad 0&0&-1&0&-1&\quad -1&2&-2&0&-1\\
 &&&\quad -1&0&0&0&1&\quad 1&0&0&0&1&\quad 1&0&0&1&0&\quad 0&0&0&0&0&\quad 1&0&0&-1&0\\
 &&&\quad -1&0&0&0&1&\quad 1&0&0&0&1&\quad -1&0&-1&-1&1&\quad 0&0&0&0&0&\quad -1&0&-1&1&1\\
 \hline
 $V_{454}^{U1}$&0.498220&8&\quad 0&0&1&0&1&\quad 0&0&1&0&1&\quad 0&0&0&0&0&\quad 0&0&0&0&0\\
 &&&\quad 1&0&0&0&1&\quad 0&-1&-1&0&0&\quad 1&-1&0&0&0&\quad 0&0&-1&0&1\\
 &&&\quad 0&-1&1&1&-1&\quad 0&-1&0&-1&0&\quad -1&1&0&0&0&\quad 1&1&-1&0&1\\
 &&&\quad -1&-1&0&1&1&\quad 0&0&0&-1&-1&\quad 0&0&0&0&0&\quad -1&-1&0&0&0\\
 \hline
 $V_{553}^{U1}$&0.496463&8&\quad -1&1&-1&-1&0&\quad -1&0&0&1&0&\quad 0&1&1&0&0\\
 &&&\quad 1&1&0&0&0&\quad 0&1&0&0&-1&\quad 1&0&0&0&1\\
 &&&\quad -1&0&1&0&0&\quad 0&0&1&0&1&\quad 1&0&0&0&1\\
 &&&\quad -1&0&0&0&1&\quad 1&0&0&0&-1&\quad 0&0&0&0&0\\
 &&&\quad 0&0&0&1&-1&\quad 0&-1&1&-1&-1&\quad 0&1&1&0&0\\
 \hline
 \end{tabular}
 \label{table:555}
 \end{table*}

 \begin{table*}[tbm]
 \caption{The Bell coefficients of $V_{555}^{2}$, the Bell inequality with five measurement settings per
 party with the critical visibility $v_{crit}=0.496059$, the second best one we found. The local bound
 is ${\cal L}=13180$.}
 \vskip 0.2truecm
 \centering
 \begin{tabular}{r r r r r r r r r r r r r r r r r r r r r r r r r}
 \hline\hline
 -271&{\hspace{-0.1 cm}}-124&{\hspace{-0.1 cm}}-418&{\hspace{-0.1 cm}}522&{\hspace{-0.1 cm}}251&{\hspace{-0.1 cm}}\quad 922&{\hspace{-0.1 cm}}355&{\hspace{-0.1 cm}}8&{\hspace{-0.1 cm}}522&{\hspace{-0.1 cm}}-37&{\hspace{-0.1 cm}}\quad 540&{\hspace{-0.1 cm}}206&{\hspace{-0.1 cm}}69&{\hspace{-0.1 cm}}90&{\hspace{-0.1 cm}}-641&{\hspace{-0.1 cm}}\quad -294&{\hspace{-0.1 cm}}437&{\hspace{-0.1 cm}}853&{\hspace{-0.1 cm}}90&{\hspace{-0.1 cm}}-212&{\hspace{-0.1 cm}}\quad 65&{\hspace{-0.1 cm}}710&{\hspace{-0.1 cm}}210&{\hspace{-0.1 cm}}0&{\hspace{-0.1 cm}}-565\\
 -514&{\hspace{-0.1 cm}}251&{\hspace{-0.1 cm}}736&{\hspace{-0.1 cm}}596&{\hspace{-0.1 cm}}-403&{\hspace{-0.1 cm}}\quad -138&{\hspace{-0.1 cm}}-45&{\hspace{-0.1 cm}}-249&{\hspace{-0.1 cm}}-612&{\hspace{-0.1 cm}}736&{\hspace{-0.1 cm}}\quad -217&{\hspace{-0.1 cm}}-641&{\hspace{-0.1 cm}}-350&{\hspace{-0.1 cm}}0&{\hspace{-0.1 cm}}932&{\hspace{-0.1 cm}}\quad 853&{\hspace{-0.1 cm}}-212&{\hspace{-0.1 cm}}-45&{\hspace{-0.1 cm}}82&{\hspace{-0.1 cm}}-514&{\hspace{-0.1 cm}}\quad 418&{\hspace{-0.1 cm}}-557&{\hspace{-0.1 cm}}-400&{\hspace{-0.1 cm}}-98&{\hspace{-0.1 cm}}-1277\\
 -446&{\hspace{-0.1 cm}}271&{\hspace{-0.1 cm}}-1146&{\hspace{-0.1 cm}}-457&{\hspace{-0.1 cm}}514&{\hspace{-0.1 cm}}\quad -636&{\hspace{-0.1 cm}}-922&{\hspace{-0.1 cm}}-482&{\hspace{-0.1 cm}}612&{\hspace{-0.1 cm}}130&{\hspace{-0.1 cm}}\quad 546&{\hspace{-0.1 cm}}-540&{\hspace{-0.1 cm}}-996&{\hspace{-0.1 cm}}-167&{\hspace{-0.1 cm}}209&{\hspace{-0.1 cm}}\quad -347&{\hspace{-0.1 cm}}294&{\hspace{-0.1 cm}}-477&{\hspace{-0.1 cm}}175&{\hspace{-0.1 cm}}-853&{\hspace{-0.1 cm}}\quad 775&{\hspace{-0.1 cm}}-65&{\hspace{-0.1 cm}}-145&{\hspace{-0.1 cm}}147&{\hspace{-0.1 cm}}-418\\
 -457&{\hspace{-0.1 cm}}-522&{\hspace{-0.1 cm}}661&{\hspace{-0.1 cm}}0&{\hspace{-0.1 cm}}-596&{\hspace{-0.1 cm}}\quad 604&{\hspace{-0.1 cm}}-514&{\hspace{-0.1 cm}}858&{\hspace{-0.1 cm}}-336&{\hspace{-0.1 cm}}612&{\hspace{-0.1 cm}}\quad -159&{\hspace{-0.1 cm}}-90&{\hspace{-0.1 cm}}-495&{\hspace{-0.1 cm}}426&{\hspace{-0.1 cm}}0&{\hspace{-0.1 cm}}\quad 159&{\hspace{-0.1 cm}}-90&{\hspace{-0.1 cm}}763&{\hspace{-0.1 cm}}604&{\hspace{-0.1 cm}}-90&{\hspace{-0.1 cm}}\quad 147&{\hspace{-0.1 cm}}-8&{\hspace{-0.1 cm}}-465&{\hspace{-0.1 cm}}-514&{\hspace{-0.1 cm}}90\\
 1146&{\hspace{-0.1 cm}}-418&{\hspace{-0.1 cm}}669&{\hspace{-0.1 cm}}-661&{\hspace{-0.1 cm}}736&{\hspace{-0.1 cm}}\quad 466&{\hspace{-0.1 cm}}8&{\hspace{-0.1 cm}}-1915&{\hspace{-0.1 cm}}-858&{\hspace{-0.1 cm}}-233&{\hspace{-0.1 cm}}\quad 996&{\hspace{-0.1 cm}}69&{\hspace{-0.1 cm}}368&{\hspace{-0.1 cm}}503&{\hspace{-0.1 cm}}-342&{\hspace{-0.1 cm}}\quad 461&{\hspace{-0.1 cm}}853&{\hspace{-0.1 cm}}342&{\hspace{-0.1 cm}}-771&{\hspace{-0.1 cm}}-37&{\hspace{-0.1 cm}}\quad 145&{\hspace{-0.1 cm}}210&{\hspace{-0.1 cm}}-510&{\hspace{-0.1 cm}}465&{\hspace{-0.1 cm}}-400\\
 \hline
 \end{tabular}
 \label{table2:555}
 \end{table*}

\section{The method}\label{sec:method}

To find Bell inequalities with low classical per quantum values
(hence low critical visibilities), we applied a two step
procedure. In this paragraph we describe the two steps briefly,
and in the next ones a more detailed description is given. In the
first step we used a simplex downhill algorithm \cite{NM} to
minimize the classical per quantum value for Bell inequalities of
very special form, characterized by just one parameter for each
measurement setting. The initial parameters were chosen randomly.
This procedure leads to non-tight Bell inequalities (that is,
which do not define facets of the local full-tripartite
correlation polytope) with not exceptionally good critical
visibilities, usually between 0.5 and 0.6. In no case we got a
desired value of less then 0.5 in this way. However, the best
critical visibility can always be achieved with a tight Bell
inequality, therefore, in the second step we determined the facet
of the polytope crossed by the ray pointing towards the direction
in the space of correlations defined by the Bell coefficients we
have got in the first step. The tight inequality corresponding to
this facet always gives a critical visibility value better than
the original inequality does. Here we worked in the restricted
space of three-party correlations, but all inequalities we have
got this way are tight inequalities in the full space as well,
which includes two-party correlation terms and single-party
marginal terms.

The coefficients of the special Bell inequalities, which multiply
the corresponding three-party correlators $A_iB_jC_k$ we
considered in the first step are given as:
\begin{equation}
M_{ijk}=\cos(\varphi^A_i+\varphi^B_j+\varphi^C_k).
\label{eq:Bellcoeff}
\end{equation}
When optimizing the critical visibility we have taken
$\sum_{i=1}^{m_A}\sum_{j=1}^{m_B}\sum_{k=1}^{m_C}M_{ijk}^2$ as the
quantum value of the Bell expression, that is the coordinates of
the point in the space of correlations for which we calculated the
quantum value are just the Bell coefficients themselves. This
point does belong to the set of quantum correlations, as it can be
achieved by von Neumann projective measurements performed on
components of a shared 3-qubit GHZ state. If we choose the $i$th,
$j$th and $k$th measurement operator $\hat A_i$, $\hat B_j$ and
$\hat C_k$ of Alice, Bob and Cecil, respectively as
\begin{align}
\hat A_i&=\cos\varphi^A_i\hat\sigma_x+\sin\varphi^A_i\hat\sigma_y\nonumber\\
\hat B_j&=\cos\varphi^B_j\hat\sigma_x+\sin\varphi^B_j\hat\sigma_y\nonumber\\
\hat C_k&=\cos\varphi^C_k\hat\sigma_x+\sin\varphi^C_k\hat\sigma_y,
\label{eq:measoper}
\end{align}
where $\hat\sigma_x$ and $\hat\sigma_y$ are Pauli operators, then
the joint correlation is given by the right hand side of
Eq.~(\ref{eq:Bellcoeff}), indeed, that is:
\begin{equation}
A_iB_jC_k=\langle\hat A_i\otimes\hat B_j\otimes\hat
C_k\rangle=\cos(\varphi^A_i+\varphi^B_j+\varphi^C_k).
\label{eq:3corr}
\end{equation}
See for example Appendix C in Ref.~\cite{BGLP} or Eqs.~(6,7) in
Ref.~\cite{PV11} for more general formulae, allowing for
non-equatorial measurements (ones with $\hat\sigma_z$ components).

We have chosen the form of the Bell coefficients to coincide with
the coordinates of an existing point in the set of quantum
correlations because for this point this is the Bell inequality
achieving the largest possible quantum value divided by the length
of the vector defined by the coefficients. The reason we allowed
only equatorial measurements to get the point in the correlation
space was that maximum quantum value for the GHZ state can always
be achieved for any full-tripartite correlation-type Bell
inequality without taking more general operators (see for instance
\cite{PV11}).

We note that the quantum value we consider as above is usually not
the maximum quantum value of the Bell inequality. Even allowing
only 3-qubit GHZ states there are usually quantum correlations
giving larger quantum violations. However, the improvement appears
usually only in the fourth or fifth significant digit by the time
the optimization procedure is finished, so it is unnecessary to
determine the true maximum violation. The objective function has
to be evaluated many times during the optimization procedure, and
just taking the sum of the squares of the Bell coefficients as the
quantum value is much faster. In principle, allowing other than
3-qubit GHZ states, possibly states in higher dimensional Hilbert
spaces might further improve the quantum value, but we have found
that for all tight Bell inequalities we arrived at in the second
step this was not the case. In the second step, similarly as in
the work of Ref.~\cite{zoo}, we use linear programming to find the
facet of the polytope in the direction of the vector defined by
the Bell coefficients we got in the first step. In this step the
critical visibility improves significantly most of the case. We
note that usually we got the best final results not from the
initial inequalities giving the lowest critical visibilities.
Those most often led to inequalities that are equivalent to the
Mermin inequality.

\section{Results}\label{sec:results}

For cases of no more than three measurement settings per party we
have found no Bell inequality with critical visibility $v_{crit}$
better than 0.5. For four measurement settings for one of the
parties and three for the other two we have found two such
inequivalent inequalities, $V_{343}^{1}$ with $v_{crit}=0.49967$,
and $V_{343}^{2}$ with $v_{crit}=0.49972$, see
Table~\ref{table:443}. We generated 30000 inequalities with the
simplex downhill method starting from different random initial
parameter values. In the second step we got inequalities
equivalent with $V_{343}^{1}$ 374 times, and with  $V_{343}^{2}$
68 times, respectively. The rest has $v_{crit}=0.5$, many of them
equivalent to the Mermin inequality, or $v_{crit}>0.5$. We have
done the same number of attempts for three measurement settings
for one of the parties and four for the other two. We arrived 1372
times at inequalities with  $v_{crit}<0.5$, among them there were
70 independent ones, one of them equivalent with $V_{343}^{2}$.
Many of the inequalities were found only once, which indicates
that there are certainly a lot we missed. The ten best ones we
have got are shown in Table~\ref{table:443}.

For four measurement settings per party we have made 10000
attempts. In about one third of these attempts we got inequalities
with $v_{crit}<0.5$, of which over one thousand were inequivalent.
The ten with the best critical visibilities are shown in
Table~\ref{table:444}. It is interesting, that the integer-valued
Bell coefficients are quite small for all of them. In
Table~\ref{table:444} we show five more inequalities with four
measurement settings per party with special properties. Two of
them have only Bell coefficients 0 and $\pm 1$, while the other
three inequalities have coefficients symmetric in the exchange of
all three parties.

We have also made 10000 calculations for five measurement settings
per party. In almost three quarters of the trials we got result
with $v_{crit}<0.5$. The best inequality, and some more
inequalities have been found several times, but we have arrived at
the majority of them only once. Some examples are shown in
Tables~\ref{table:555} and \ref{table2:555}. The improvement in
$v_{crit}$ compared to the four measurement settings per party
case is not very impressive. We show the three best cases, the
second best in Table~\ref{table2:555} separately, for typographic
reason: the integer Bell coefficients are fairly large for that
case. In Table~\ref{table:555} we also show some special cases.
$V_{555}^{S1}$ is symmetric in all participants (the analytical
solution of this inequality will be given in
Sec.~\ref{sec:analytic}), while $V_{555}^{U1}$ and $V_{555}^{U2}$
have only coefficients 0 and $\pm 1$. In a few cases we got
inequalities equivalent to ones with less than five measurement
settings per party. One example is $V_{455}^{1}$, with only four
measurement settings for Alice, for which $v_{crit}$ is only very
marginally worse than for $V_{555}^{1}$. $V_{454}^{U1}$ has only 0
and $\pm 1$ Bell coefficients and two participants have 4
measurement settings, only one has five. The total number of
measurement settings is also 13 for $V_{553}^{U1}$, which also has
Bell coefficients of absolute values at most one, and which is
symmetric in the coefficients of Alice and Bob, who have five
measurement settings each, while Cecil has only three. For this
case $v_{crit}$ is quite small as well.

For the critical visibility, we calculated the quantum bound with
von Neumann projective measurements performed on components of a
shared 3-qubit GHZ state with a see-saw method, similar to
Ref.~\cite{seesaw}. For all cases such calculation has given the
exact quantum bound. This is true not only for the cases we
reported here, but for all tight three-party full-correlation type
Bell inequalities (nonzero coefficients only for three party
correlators) we checked, including the ones with worse critical
visibilities. We checked all cases we got with no more than four
measurement settings per party, and also several larger cases,
including ones with five measurement settings per party. We have
done that by calculating the upper bound for the quantum value
using semidefinite programming with the method invented by
Navascu\'es, Pironio and Ac\'in (NPA) \cite{NPA1}. For the
calculation we used the code CSDP of Brian Borchers \cite{CSDP}.
In all cases we found the upper bound coincides with the result of
direct calculation, which is actually a lower bound. In applying
the NPA method, it was enough to do it at level one
$+ab+ac+bc+abc$ in all cases. The notion of levels, and their
notation is explained in Ref.~\cite{NPA2}, and also in
Ref~\cite{PV09}.

\section{Study of a Bell inequality providing visibility less than $1/2$.}\label{sec:analytic}

We consider a Bell scenario where each of the three parties can
choose between five possible binary measurements. We denote the
outcome of measurement $j=0,1,\ldots,4$ for Alice by $A_j$ ($B_j$
and $C_j$ for Bob and Cecil respectively). Here we shall focus on
the only symmetric inequality we found for this scenario (and can
be found in Table~\ref{table:555}), which is given by the
expression:
\begin{align}
V_{555}^{S1}=&\text{sym}[- A_0B_0C_2 - A_0B_0C_3 + A_0B_2C_4 \nonumber\\
&+ A_0B_3C_4 + A_1B_1C_2 - A_1B_1C_3 - A_1B_2C_2  \nonumber\\
&+ A_1B_3C_3- 2A_2B_2C_2 + A_2B_2C_3 - A_2B_3C_3 \nonumber\\
&+ A_2B_4C_4 + 2A_3B_3C_3 + A_3B_4C_4],
\label{555sym}
\end{align}
where the designation $\text{sym}[X]$ indicates that the expression $X$
is symmetrized over all the parties, for instance
$\text{sym}[A_0B_0C_2]=A_0B_0C_2+A_0B_2C_0+A_2B_0C_0$. In this
$5\times 5\times 5$ Bell inequality the local bound is ${\cal
L}=12$ (all local probability distributions satisfy the inequality
$V_{555}^{S1}\leq 12$).

We now show that by performing local measurements on a 3-qubit GHZ state,
\begin{equation}
|GHZ\rangle=(|000\rangle+|111\rangle)/\sqrt 2,
\end{equation}
it is possible to obtain the value of ${\cal Q}=24.1699$ for the
Bell expression~(\ref{555sym}) above. This implies the visibility
of $v_{crit}=12/24.1699=0.496485$ showed in Table~\ref{table:555}.

The local measurements used by each party are of the simple form
$\hat A_j = \cos{\varphi_j}\hat\sigma_x +
\sin{\varphi_j}\hat\sigma_y$ (with $j=0, 1, \ldots, 4$), where
$\hat\sigma_x$ and $\hat\sigma_y$ are Pauli matrices. Moreover,
the five local measurements are the same for each party, i.e.
$\hat A_j=\hat B_j=\hat C_j$, $j=0,1,2,3,4$. In this case the
3-party correlation term has the following form (see
Eq.~(\ref{eq:3corr})),
\begin{equation}
A_iB_jC_k=\cos(\varphi_i+\varphi_j+\varphi_k).
\end{equation}
The optimal measurements are given by $\varphi_0=5\pi/8$,
$\varphi_1=\pi/2$, $\varphi_4=\pi/8$, and
$\varphi_2+\varphi_3=3\pi$. Leaving $\varphi_2$ as a free
parameter, we obtain the following closed formula for the Bell
value,
\begin{equation}
S=-12\cos3\varphi_2+12\cos\varphi_2(\sin\varphi_2-1)-12\sqrt
2\sin\varphi_2.
\label{eq:1param}
\end{equation}
Our task is now to perform optimization over the angle
$\varphi_2$. By inserting $\varphi_2=3.73842$
into~(\ref{eq:1param}), we obtain the value of ${\cal Q}=24.1699$
for the expression~(\ref{eq:1param}).

\section{Conclusion}\label{sec:conc}

The non-local properties of the noisy three-qubit
Greenberger-Horne-Zelinger (GHZ) states (defined by Eq.~(\ref{eq:noisyGHZ})),
parameterized by the visibility $0\le v\le 1$ were investigated.
Based on the violation of the $2\times 2\times 2$-setting Mermin inequality,
$\rho_v$ is non-local for the parameter range $1/2<v\le 1$. In this study we
presented several Bell inequalities which beat the limit of
$v=1/2$, thereby answering the question raised by Kaszlikowski et
al.~\cite{kz}. In particular, the lowest threshold visibility we found is
$v=0.496057$, attainable with $5\times 5\times 5$ settings, whereas
the most economical one overcoming $v=1/2$ corresponds to $3\times
3\times 4$ settings. The method which enabled us to obtain these
results, and in particular the about 10000 tight Bell inequalities
going below $v=1/2$ were also discussed in detail.

We expect that the presented method in Sec.~\ref{sec:method} would
be efficient to test the non-locality of other states as well. For
instance, in the two-party 2-qubit scenario it might prove useful
to find Werner states with better visibility thresholds than the
presently known ones \cite{ver08} (or to find ones which overcome
the $1/\sqrt 2$ limit corresponding to the CHSH inequality with
fewer number of settings). Our technique may also be suitable to
generate Bell inequalities for 3-qubit W states \cite{DVC} or
4-qubit Dicke states \cite{dicke}, cluster states \cite{cluster}
with improved noise resistance.

Further, it would be interesting to study whether the technique
presented could be applied to find Bell violation with lower
detection efficiencies than the presently known ones both for
bipartite and multipartite settings. In the bipartite case,
according to Ref.~\cite{VPB}, one may postulate the $d\times d$
state to be $|\psi\rangle=\epsilon|00\rangle + |11\rangle + \ldots
+ |d-1,d-1\rangle$ (here the state is written in an unnormalized
form), $\epsilon$ is a small parameter and the measurements are
rank-1 projectors. Then, by using appropriate parametrization, one
might carry out the heuristic search of Sec.~\ref{sec:method} for
2-outcome Bell inequalities to lower the detection efficiency
required for closing the detection loophole.

In the tripartite case, it is known that for the GHZ state the
minimum detection efficiency based on the Mermin inequality is
$75\%$ \cite{detloophole}. On the other hand, the best protocol
which simulates GHZ correlations (arising from equatorial von
Neumann measurements) with detection efficiencies of $50\%$ is due
to the recent work of Ref.~\cite{BG}. It would be interesting to
decrease the gap by considering inequalities with more than 2
settings \cite{stefano} in conjunction with the technique used in
this paper.

\emph{Acknowledgements.} T.V. acknowledges financial support from
the J\'anos Bolyai Programme of the Hungarian Academy of Sciences.


\begin{thebibliography}{99}

\bibitem{horo}
R. Horodecki et al., Rev. Mod. Phys. {\bf 81}, 865 (2009).

\bibitem{bell}
J.S. Bell, Physics {\bf 1} (1964), 195.

\bibitem{ekert} A.K. Ekert, Phys. Rev. Lett. {\bf 67},  661  (1991).

\bibitem{secure}
A. Ac\'in, N. Brunner, N. Gisin, S. Massar, S. Pironio, and V.
Scarani, Phys. Rev. Lett. {\bf 98}, 230501 (2007).

\bibitem{randomness}
S. Pironio et al., Nature {\bf 464}, 1021 (2010).

\bibitem{SG}
V. Scarani and N. Gisin, Phys. Rev. Lett. {\bf 87}, 117901 (2001).

\bibitem{werner} R.F. Werner, Phys. Rev. A {\bf 40}, 4277 (1989).

\bibitem{CHSH}
J.F. Clauser, M.A. Horne, A. Shimony and R.A. Holt, Phys. Rev. Lett. {\bf 23}, 880 (1969).

\bibitem{zuk99} M. \.Zukowski, D. Kaszlikowski, A. Baturo, and J.-A. Larsson, arxiv:quant-ph/9910058 (1999).

\bibitem{ver08} T. V\'ertesi, Phys. Rev. A {\bf 78}, 032112 (2008).

\bibitem{mermin}
N.D. Mermin, Phys. Rev. Lett. {\bf 65}, 1838 (1990); M. Ardehali, Phys. Rev. A {\bf 46}, 5375 (1992); A.V. Belinskii and D.N. Klyshko, Phys. Usp. {\bf 36}, 653 (1993).

\bibitem{DCT} W. D\"ur, J.I. Cirac, and R. Tarrach, Phys. Rev. Lett. {\bf 83}, 3562 (1999).

\bibitem{kz}
D. Kaszlikowski and M. \.Zukowski, Int. J. Theor. Phys. {\bf 42},
1023 (2003); arXiv:quant-ph/0302165 (2003).

\bibitem{gruca}
J.~Gruca, W.~Laskowski, M.~\.Zukowski, N.~Kiesel, W.~Wieczorek,
C.~Schmid, H.~Weinfurter, Phys. Rev. A {\bf 82}, 012118 (2010).

\bibitem{simul}
A. Broadbent, P.-R. Chouha, and A. Tapp, Third International
Conference on Quantum, Nano, and Micro Technologies pp. 59–62 (2009);
J.-D. Bancal, C. Branciard, and N. Gisin, Adv. Math. Phys. Article ID 293245 (2010);
C. Palazuelos, D. Perez-Garcia, and I. Villanueva, arXiv:1006.5318 (2010);

\bibitem{BG}
C. Branciard, N. Gisin, Phys. Rev. Lett. {\bf 107}, 020401 (2011).

\bibitem{WW}
R.F. Werner and M. Wolf, Phys. Rev. A {\bf 64}, 032112 (2001).

\bibitem{sliwa}
C. \'{S}liwa, Phys. Lett. A {\bf 317}, 165 (2003).

\bibitem{GHZ}
D.M. Greenberger, M.A. Horne, and A. Zeilinger, {\it Bells Theorem,
Quantum Theory, and Conceptions of the Universe} (ed. M. Kafatos,
Kluwer Academic, Dordrecht, Holland, 1989), pp. 69–-72.

\bibitem{BBT}
G. Brassard, A. Broadbent, A. Tapp, Foundations of
Physics {\bf 35}, 1877 (2005).

\bibitem{NM}
J.A. Nelder and R. Mead, Comput. J. {\bf 7}, 308 (1965).

\bibitem{BGLP}
J.-D. Bancal, N. Gisin, Y.-C. Liang, and S. Pironio, Phys. Rev. Lett. {\bf 106}, 250404 (2011).

\bibitem{PV11}
K.F. P\'al and T. V\'ertesi, Phys. Rev. A {\bf 83}, 062123 (2011).

\bibitem{seesaw} K.F. P\'al and T. V\'ertesi, Phys. Rev. A {\bf 82}, 022116 (2010).

\bibitem{zoo}
S. Massar, S. Pironio, J. Roland, and B. Gisin, Phys. Rev. A {\bf
66}, 052112 (2002).

\bibitem{NPA1}
M. Navascu\'es, S. Pironio, and A. Ac\'in, Phys. Rev. Lett. {\bf 98}, 010401 (2007).

\bibitem{CSDP}
B.~Borchers, Optimization Methods and Software {\bf 11}, 597, (1999);
B.~Borchers, Optimization Methods and Software {\bf 11}, 613, (1999);
B.~Borchers and J.G.~Young, Computational Optimization and Applications {\bf 37}, 355 (2007); \verb+https://projects.coin-or.org/Csdp/+.

\bibitem{NPA2}
M. Navascu\'es, S. Pironio, and A. Ac\'in, New. J. Phys. {\bf 10}, 073013 (2008).

\bibitem{PV09}
K.F. P\'al and T. V\'ertesi, Phys. Rev. A {\bf 79}, 022120 (2009).

\bibitem{DVC}
W. D\"ur, G. Vidal, and J.I. Cirac, Phys. Rev. A {\bf 62}, 062314 (2000).

\bibitem{dicke}
R.H. Dicke, Phys. Rev. A {\bf 93}, 99 (1954).

\bibitem{cluster}
H.J. Briegel and R. Raussendorf, Phys. Rev. Lett. {\bf 86}, 910 (2001).

\bibitem{VPB}
T. V\'ertesi, S. Pironio, and N. Brunner, Phys. Rev. Lett. {\bf 104}, 060401 (2010).

\bibitem{detloophole}
J.-A. Larsson, Phys. Rev. A {\bf 57}, R3145 (1998); ibid {\bf 59}, 4801
(1999). A. Cabello, D. Rodriguez, I. Villanueva, Phys. Rev. Lett. {\bf 101} 120402 (2008).

\bibitem{stefano}
S. Pironio, private communication.

\end{thebibliography}
\end{document}